\documentclass[iop, apj, revtex4]{emulateapj}
\usepackage{natbib}
\usepackage{url}
\usepackage{amsmath}
\usepackage{appendix}
\usepackage{hyperref}
\begin{document}

\title{Thick Disks in the Hubble Space Telescope Frontier Fields}

\author{Bruce G. Elmegreen\altaffilmark{1}, Debra Meloy Elmegreen\altaffilmark{2},
Brittany Tompkins\altaffilmark{2}, Leah G. Jenks\altaffilmark{2,3}}

\altaffiltext{1}{IBM Research Division, T.J. Watson Research Center, 1101 Kitchawan
Road, Yorktown Heights, NY 10598; bge@us.ibm.com}

\altaffiltext{2}{Department of Physics \& Astronomy, Vassar College, Poughkeepsie,
NY 12604; elmegreen@vassar.edu}

\altaffiltext{3}{Colgate University, Department of Physics and Astronomy, Hamilton, NY
13346}

\begin{abstract}
Thick disk evolution is studied using edge-on galaxies in two Hubble Space Telescope
Frontier Field Parallels. The galaxies were separated into 72 clumpy types and 35
spiral types with bulges. Perpendicular light profiles in F435W, F606W and F814W (B, V
and I) passbands were measured at 1 pixel intervals along the major axes and fitted to
${\rm sech}^2$ functions convolved with the instrument line spread function (LSF).  The
LSF was determined from the average point spread function (PSF) of $\sim20$ stars in
each passband and field, convolved with a line of uniform brightness to simulate disk
blurring.  A spread function for a clumpy disk was also used for comparison. The
resulting scale heights were found to be proportional to galactic mass, with the
average height for a $10^{10\pm0.5}\;M_\odot$ galaxy at $z=2\pm0.5$ equal to
$0.63\pm0.24$ kpc. This value is probably the result of a blend between thin and thick
disk components that cannot be resolved. Evidence for such two-component structure is
present in an inverse correlation between height and midplane surface brightness.
Models suggest that the thick disk is observed best between the clumps, and there the
average scale height is $1.06\pm0.43$ kpc for the same mass and redshift. A
$0.63\pm0.68$ mag V-I color differential with height is also evidence for a mixture of
thin and thick components.
\end{abstract}


\section{Introduction}
\label{intro}

Thick disks are an old and faint component of most modern galaxies
\citep{burstein79,gilmore83,dalcanton02,yoachim06,yoachim08b,comeron11,comeron14}. They
have high $\alpha/Fe$ abundance ratios, indicating rapid formation, and they have low
metallicities and red colors because they are old \citep[e.g.,][]{fuhrmann98,
bensby07}. They were much brighter in the early universe when they were young, as can
be seen directly in Hubble Space Telescope deep fields where the angular resolution is
typically comparable to the thick-disk scale height \citep{elmegreen06}.

Observations of the thick and thin disks in the Milky Way suggest that the transition
from one to the other over time was relatively smooth. There is a continuous change in
the distribution of stellar mass \citep{bovy12a} and height \citep{bovy12b} as stars
change from the thick disk with high $\alpha/Fe$ and low $Fe/H$ to the thin disk. The
distribution of stellar surface density as a function of scale height is also smooth,
with a steady increase in surface density while the scale height decreases
\citep{bovy12a}.

A sharper difference between Milky Way thick and thin disk stars is shown by the
distribution of $\alpha/Fe$ versus $Fe/H$ for different intervals of height and
galactocentric radius \citep{hayden15}. High $\alpha/Fe$ stars show up primarily at
heights $z>1-2$ kpc and radii $R<9$ kpc, while low $\alpha/Fe$ stars occupy large
heights, $z>1-2$ kpc, primarily in the outer disk, $R>9$ kpc, and low heights
throughout the whole disk. There is also a distinction in the distribution of azimuthal
speed, $V_{\phi}$, versus $Fe/H$ \citep{lee11}: for the thin disk, $V_{\phi}$ decreases
with increasing $Fe/H$ to values even larger than solar, reflecting the selection of
inner-disk stars near the Sun at the outer parts of their epicycles and outer-disk
stars near the Sun at the inner parts of their epicycles. On the other hand, for the
thick disk, $V_{\phi}$ increases with increasing $Fe/H$, up to and slightly overlapping
with the lower limit on $Fe/H$ for the thin disk, reflecting an increasing rotation
speed for stars orbiting closer to the midplane. The latter is the effect of stellar
pressure on orbit speed, i.e., asymmetric drift \citep{binney08}.  \cite{liu12} also
found that thick disk stars have higher eccentricities than thin disk stars, consistent
with an early phase of scattering by massive clumps \citep{bournaud09}, although
simulations of this process by \cite{inoue14} did not reproduce the Milky Way's
decrease in eccentricity for increasing metallicity.

The Milky Way thick disk is apparently shorter than the thin disk
\citep{bovy12c,cheng12,bensby11,bensby17}, reflecting inside-out growth from an
increase in accreted angular momentum over time \citep{pichon11}. This change is
reproduced in Milky Way simulations, where the disk scale length increases as the scale
height decreases in a continuous fashion
\citep{sanchezb09,bird13,stinson13,aumer13,martig14,minchev15,athanassoula16}. If a
galaxy forms quickly, such as a massive disk galaxy \citep{behroozi13}, then both the
thick (early-forming) and thin (late-forming) components can be old today, as observed
by \cite{comeron16} and \cite{kasparova16}.  At the other extreme, dwarf Irregular
galaxies form slowly and are nearly pure thick disks today in terms of the ratio of
height to radius \citep[e.g.,][]{eh15,comeron14}.  Because the star formation rate in
galaxies depends on their mass \citep{behroozi13}, galaxies more massive than the Milky
Way should also have high $\alpha/Fe$ in both their thick disks and the older parts of
their thin disks.  Disks less massive than the Milky Way should have low $\alpha/Fe$
even in their thick parts.

Thick disks are expected for young galaxies because the turbulent speed in the gas is
high compared to the rotation speed \citep{forster09,kassin12}. Then stars that form in
this gas will be relatively high-dispersion too.  High turbulent speeds may result from
rapid accretion, disk instabilities, and stellar feedback
\citep{bournaud09,eb10,martig14,forbes14}. Observations in support of this turbulent
picture are in \cite{robin14}; \cite{aumer16} show that stellar scattering from giant
molecular clouds is not enough to make the thick disk. Thick disks can also form from
thin disks that have a minor merger, but then the thick disks might be expected to
flare and that is not observed \citep{bournaud09,comeron11,robin14}. Also with mergers,
some examples should be seen with counter-rotation. \cite{comeron15} state that in the
one case they observed, the thick disk rotates in the prograde direction, while
\cite{yoachim08a} report a possible case with counter-rotation. Another model is that
stellar migration in a thin disk gives the appearance of a thick disk component, but
\cite{veraciro14} and \cite{veraciro16} suggest that migration affects only the thin
disk.

This paper measures the disk thicknesses of galaxies in the HST Frontier Fields
\citep{lotz17} Abell 2744 Parallel and MACS J0416.1-2403 Parallel (hereafter, galaxies
from these fields are abbreviated Axxxx and Mxxxx), which have tabulated redshifts and
masses. The sample and instrument corrections are discussed in Section 2 and the
results are in Section 3.

\section{Data and Methods}

\subsection{Sample Selection}

Hubble Space Telescope archival data of two Frontier Fields were used for this study.
The Frontier Fields are a set of deep images that consist of six galaxy clusters and
six corresponding parallel fields. Abell 2744 and MACS J0416.1-2403 are the first
fields that were catalogued and publicly available; here we examine their corresponding
parallel fields. For our analysis we utilized images in the F435W, F606W and F814W
filters, hereafter abbreviated as B, V and I. These were taken by the Hubble ACS camera
and have an image depth of 140 orbits. The images are 10,800 x 10,800 pixels and have a
scale of 0.03 arcsec pixel$^{-1}$. Data on the fields were obtained from the AstroDeep
Frontier Fields Catalogues, given by \cite{castellano16} and \cite{merlin16}. These
catalogues were compiled using multi-wavelength photometry and spectral energy
distributions to determine redshifts and restframe galaxy properties. With IRAF (Image
Reduction and Analysis Facility) and DS9, we identified by eye and classified all
galaxies in the two fields that appeared to be linear and edge-on, resulting in a first
sample of 188 galaxies.

Among the identified galaxies, 44 were excluded from further analysis due to several
causes: faintness, a possible merger remnant, not edge-on or not straight, or having
redshifts $\sim0.01$ in the catalog, which is inconsistent with nearby high-redshift
galaxies that look similar. The remaining 144 galaxies from both fields were analyzed.
These had angular radii larger than 10 pixels (0.3 arcsec) and physical radii larger
than $\sim2$ kpc. The outer reliable isophotes of the galaxies correspond to $\sim28$
mag arcsec$^{-2}$ in surface brightness. Most of the selected galaxies fall into the
redshift range of z=0.5 to 4.

Galaxies were first classified into spiral, clumpy, spheroidal and transition types
using the images in DS9 and midplane light profiles from the task {\it pvector} in
IRAF. The clumpy galaxies are characterized by distinct regions of star formation,
visible as clumps in the images and as peaks in the radial profiles of these galaxies.
Spiral galaxies were identified by a central bulge and an exponential radial profile.
Transition galaxies did not fit cleanly into the clumpy or spiral categories based on
their radial profiles, but often showed a bulge with a relatively large second clump.
Spheroidal galaxies did not show observable clumps or exponential radial profiles, and
had indistinct structure. Our sample had 72 clumpy galaxies, 35 spiral galaxies, 30
transition galaxies, and 7 spheroidal galaxies in the two fields combined. After
further consideration, we eliminated the transition and spheroidal types as we could
not be certain that the associated perpendicular intensity profiles were really
sampling the thicknesses of the stellar distributions rather than a warp or tidal
feature. Here we discuss the perpendicular profiles of the 72 clumpy galaxies and 35
spirals.

\subsection{Point Spread Function for Image Deconvolution}
\label{sect:lsf}

The point spread function of the instrument was evaluated by stacking approximately 20
stars in each field in B, V, and I passbands. Four scans spaced by equal angles through
the stacked stellar images were then used to find the average stellar profiles. The
stacked image looked relatively symmetric and the four scans were similar to each
other.

\begin{figure}
\includegraphics[scale=0.5]{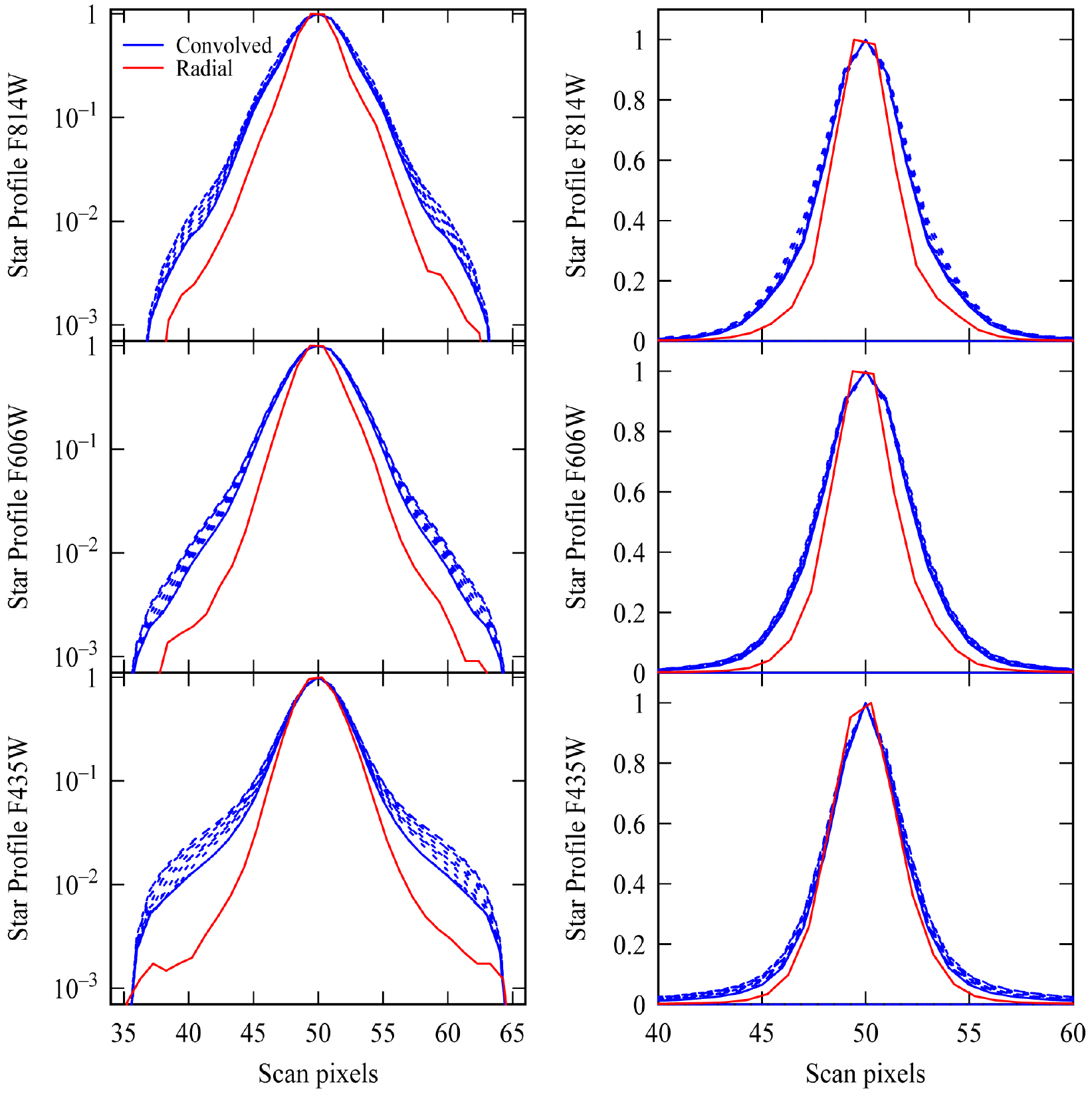}
\caption{The average radial scans through stacked stellar images are shown
as red curves for the three passbands of our observations.
The convolutions of the two-dimensional
stellar images with a line of brightness, the line spread functions (LSF),
are shown as solid-line blue curves. The profiles are
the same on the left and right, with the left using a log scale in the vertical direction to
highlight the slight differences between the stellar images and the LSF at low intensities.
Dotted blue curves use a clumpy LSF.
} \label{frontier_checkstarprofile}
\end{figure}

Image deconvolution by a symmetric PSF depends on the shape of the imaged source. If
the source is another star, then its deconvolution can be done using only the average
linear scan through the PSF. However, if the source is an edge-on galaxy disk, then
deconvolution can only be done with a ``line spread function'' (LSF), which is the
convolution of the PSF with a line of infinitesimal thickness.  The LSFs for each
passband in field Abell 2744 are shown in Figure \ref{frontier_checkstarprofile} (blue
solid curves), along with the intensity scans through the stacked stars (red curves).
The left-hand side uses a logarithmic scale to highlight the differences in the profile
wings; the right-hand side shows the same profiles with a linear scale. The LSF is
broader than the PSF below $\sim10$\% of the peak because the PSF has broad
non-Gaussian wings. If the PSF were point-symmetric and Gaussian, then the LSF would be
the same as the PSF because integrations through a Gaussian in two orthogonal
directions are independent. The profiles for the other field, MACS J0416.1-2403, are
similar and not shown. See also \cite{comeron17} for a discussion of this modification
to the PSF for edge-on galaxies. \cite{sandin15} discusses possible problems in the
measurement of thick disks because of improper removal of the PSF.

A clumpy line spread function (CLSF) was also made by convolving the stellar PSF with a
line of brightness on which is superposed a uniformly bright region 6 pixels wide (the
``clump''), which is the typical width of a large clump, and 6 times the line
intensity, which is the typical brightness factor. The reason for doing this was to
assess the effect of clumps on perpendicular scans through the interclump regions; some
of that clump emission can contaminate the interclump scan because of its presence in
the wings of the PSF. We found that when the bright part of the CLSF passes through the
center of the PSF (as if the perpendicular intensity scan were going through a clump),
the CLSF looks essentially the same as the LSF because both the clump and the line are
dominated by the sharp central part of the PSF peak. When the bright part of the CLSF
is far from the center of the PSF, the CLSF is again the same as the LSF because the
PSF is too weak at this distance to pick up the clump. At intermediate distances, the
bright part adds to the wings of the CLSF and makes it a little broader than the pure
LSF. The blue dotted curves in Figure \ref{frontier_checkstarprofile} show these CLSF
for clump central positions starting at 3 pixels from the peak of the PSF and with
uniform pixel spacings up to 13 pixels from the peak. The excess over the LSF increases
with distance at first and then decreases. The maximum excess for the F814W passband
occurs around 8 pixels from the PSF center (Fig. \ref{frontier_checkstarprofile}) and
equals a factor of 3.6 for a clump center that is 4 to 5 pixels from the center. This
excess factor does not affect the results much (see below) because it is in a part of
the LSF that is already down from the peak by a factor of $\sim100$.

\section{Results}

\subsection{Scale Height Determination and an anti-correlation with Intensity}
\label{anti}

\begin{figure}
\includegraphics[scale=0.3]{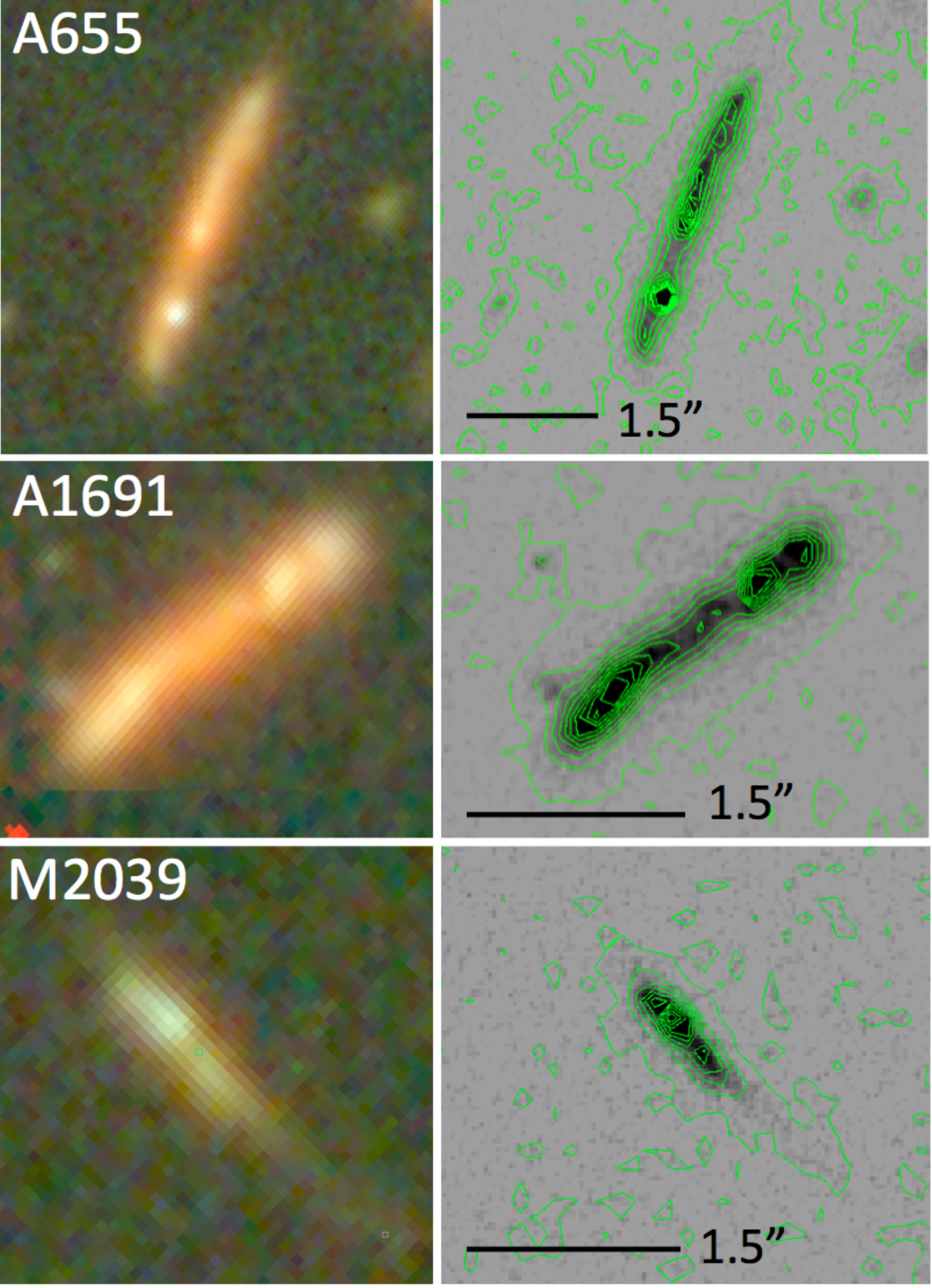}
\caption{Three clumpy galaxies are shown in color on the left using
F435W, F814W, and F160W filters.  Contours for the same galaxies are shown on the right
using contour values ranging from 0 to 0.01 counts in steps of 0.001 counts.
The contours at the apparent edge of the optical image are around $3\sigma$.}
\label{3clumpies-contours}
\end{figure}

\begin{figure}
\includegraphics[scale=0.4]{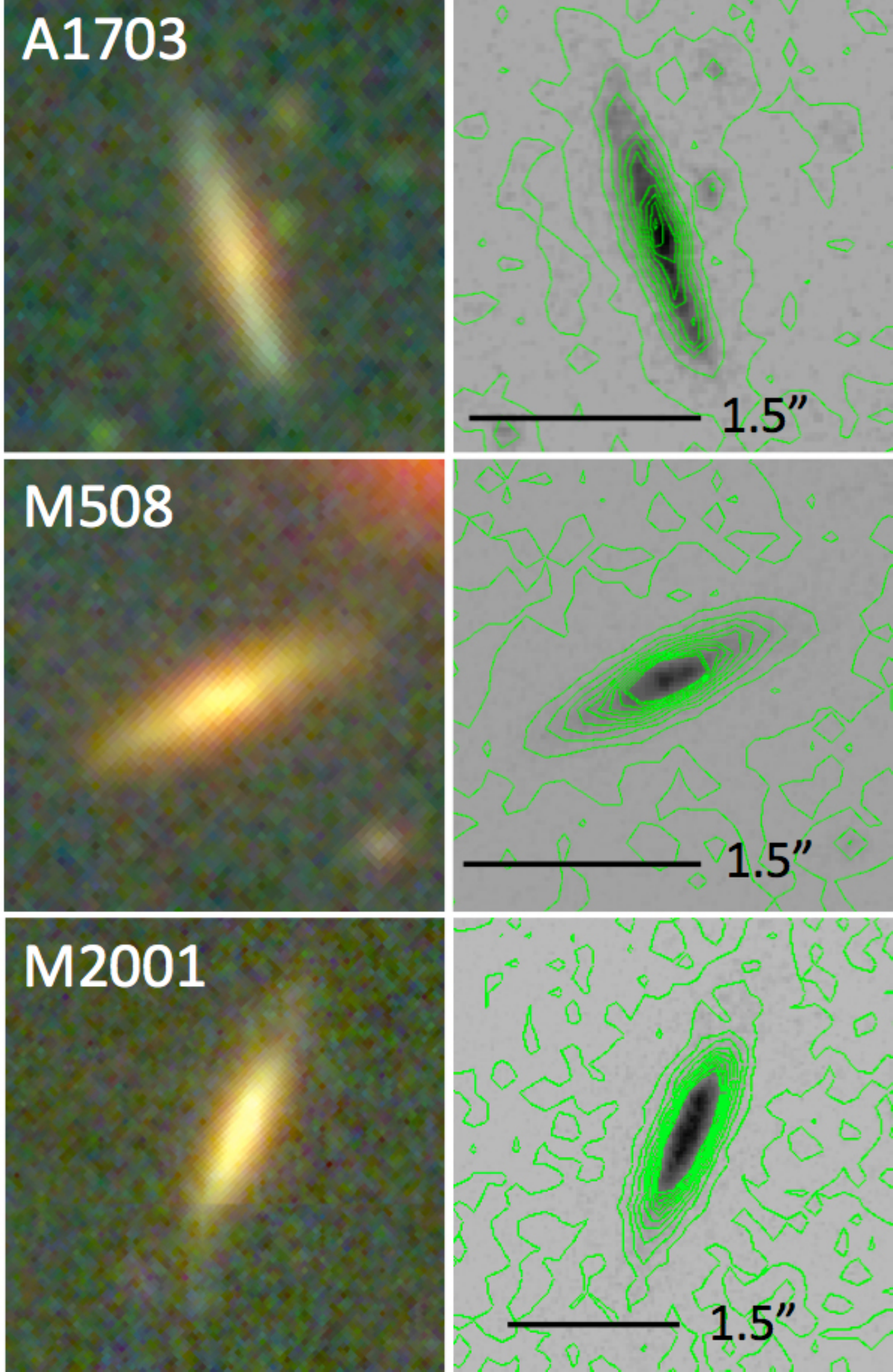}
\caption{Three spiral galaxies are shown in color on the left using
F435W, F814W, and F160W filters.  Contours are on the right
ranging from 0 to 0.01 counts in steps of 0.001 counts. } \label{3spirals-contours}
\end{figure}

Figures \ref{3clumpies-contours} and \ref{3spirals-contours} show three clumpy galaxies
and three spiral galaxies along with their surface brightness contours. Intensity
contours are from 0 to 0.01 counts in steps of 0.001 counts. The contours at the
apparent edge of the optical image are $3\sigma$.  These $3\sigma$ contours are fairly
straight around the clumpy galaxies, and they are bowed-out around the bulge regions of
the spiral galaxies. We discuss in Section \ref{model} the implication of this
observation, which is that the perpendicular scale height tends to decrease near the
clumps in clumpy galaxies, while it is more constant throughout the bulge and disk of
spiral galaxies. First, this trend will be shown for all of the galaxies here.

\begin{figure}
\includegraphics[scale=1.1]{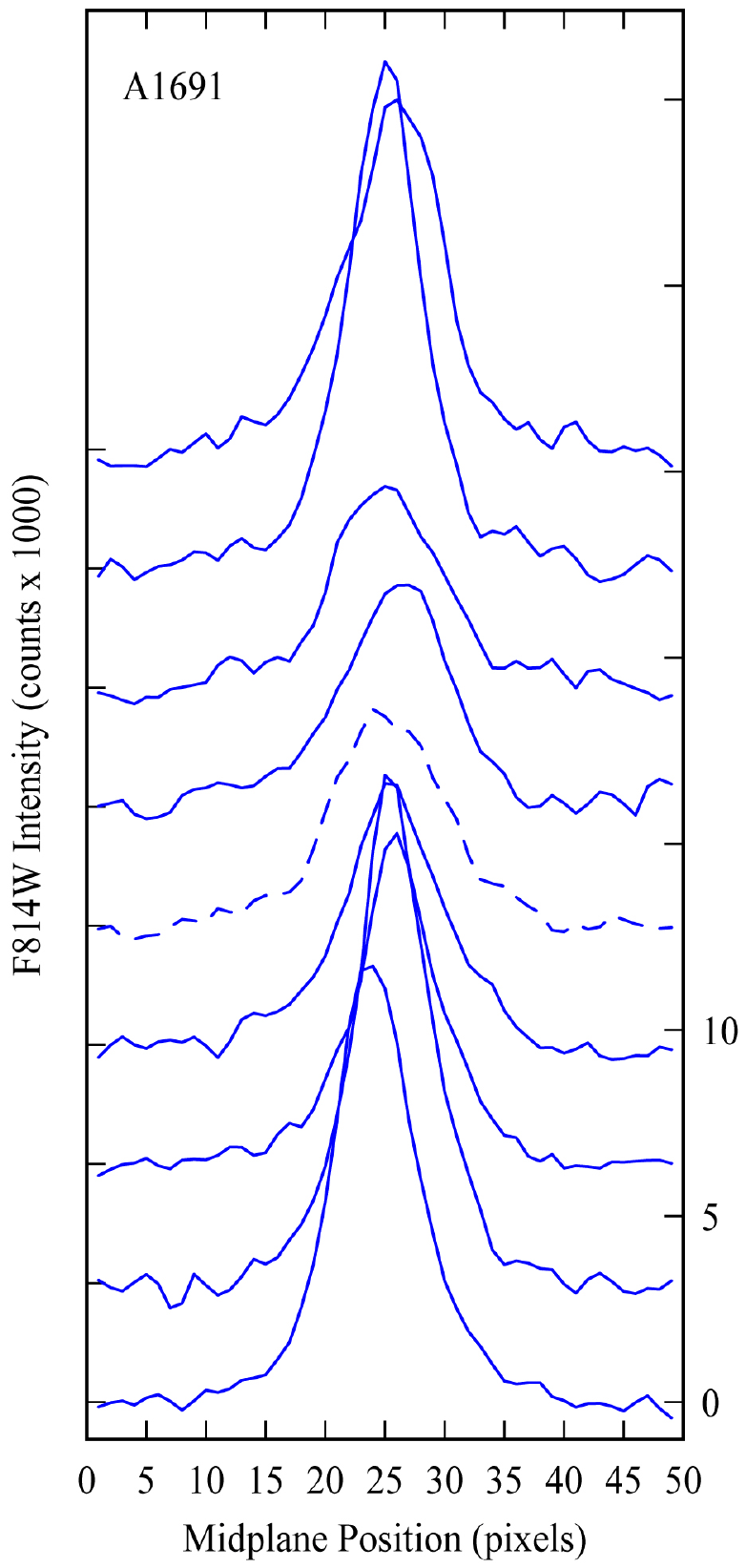}
\caption{Intensity scans perpendicular to the disk of the clumpy galaxy
A1691 taken every 5 pixels along the midplane. The central scan at pixel number 40 is
indicated by a dashed line.} \label{frontier_allprofiles_1691}
\end{figure}

Figure \ref{frontier_allprofiles_1691} shows perpendicular I-band intensity profiles in
steps of 5 pixels along the major axis of A1691, which is the middle galaxy in Figure
\ref{3clumpies-contours}. The intensity, measured in image counts$\times1000$, peaks at
the center of each scan, which is the midplane of the disk. The peaks are higher near
the ends of the galaxy because of bright clumps (Fig. \ref{3clumpies-contours}).  The
scan near the center of the galaxy, which is pixel number 40, is shown as a dashed
line.

Perpendicular intensity scans like these were measured for each galaxy and passband and
spaced by single pixel increments along the major axes. The scans were fitted to ${\rm
sech}^{2}$ functions that were blurred by the line spread function, LSF, discussed in
Section 2. To be specific, the fitting minimized the rms difference between a blurred
${\rm sech}^{2}([i-i_0]/H)$ function of the pixel number, $i$, and the intensity scan
normalized to unit height, $I_{\rm norm}(i)$, with fitting parameters equal to the
center of the model function, $i_0$, in pixels, and the width of the model function,
$H$, also in pixels. In other words, we determined the function
\begin{equation}
{\rm sech}^{2}([i-i_0]/H)=\left({{2}\over{e^{(i-i_0)/H}+e^{-(i-i_0)/H}}}\right)^2
\end{equation}
such that the convolution, $C(i)$, of ${\rm sech}^2$ with the LSF, $L(i)$, i.e.,
\begin{equation}
C(i) = \Sigma_{i^\prime=-\infty}^{\infty} L(i-i^\prime){\rm sech}^2([i^\prime-i_0]/H)di^\prime
\end{equation}
has the smallest rms compared to the observations, $I_{\rm norm}(i)$, when $C(i)$ is
also normalized,
\begin{equation}
{\rm rms}^2 = \Sigma_{i=-\infty}^{\infty} \left(I_{\rm norm}[i]-C_{\rm norm}[i]\right)^2 di.
\end{equation}
This fitting was done numerically by iteration.

\begin{figure}
\includegraphics[scale=1.]{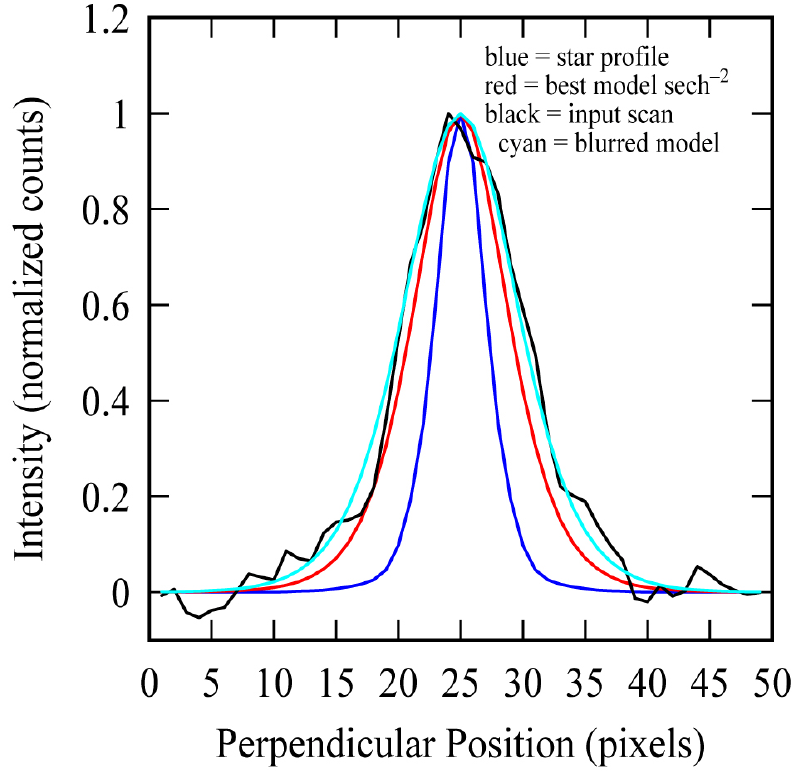}
\caption{Sample F814W profile through the center of the clumpy galaxy A1691 (black curve)
with the best-fit, LSF-convolved profile shown in cyan, along with the intrinsic
disk profile before convolution with the LSF in red, and the stellar LSF in blue. }
\label{frontier_profiles_5-28-17}
\end{figure}

Figure \ref{frontier_profiles_5-28-17} shows a typical result from the perpendicular
profile at position 40 in the galaxy A1691, which is the dashed profile in Figure
\ref{frontier_allprofiles_1691}. The black curve is the measured profile (``input
scan'' in the figure label). The blue curve is the stellar LSF, the red curve is the
fitted ${\rm sech}^{2}$ function, and the cyan curve is the LSF-blurred ${\rm
sech}^{2}$ function, which is the best match to the observations. All of the profiles
have been normalized to unit height. In this case, $H=5.0$ pixels and the rms deviation
for the fit is $0.038$ counts.

\begin{figure}
\includegraphics[scale=0.5]{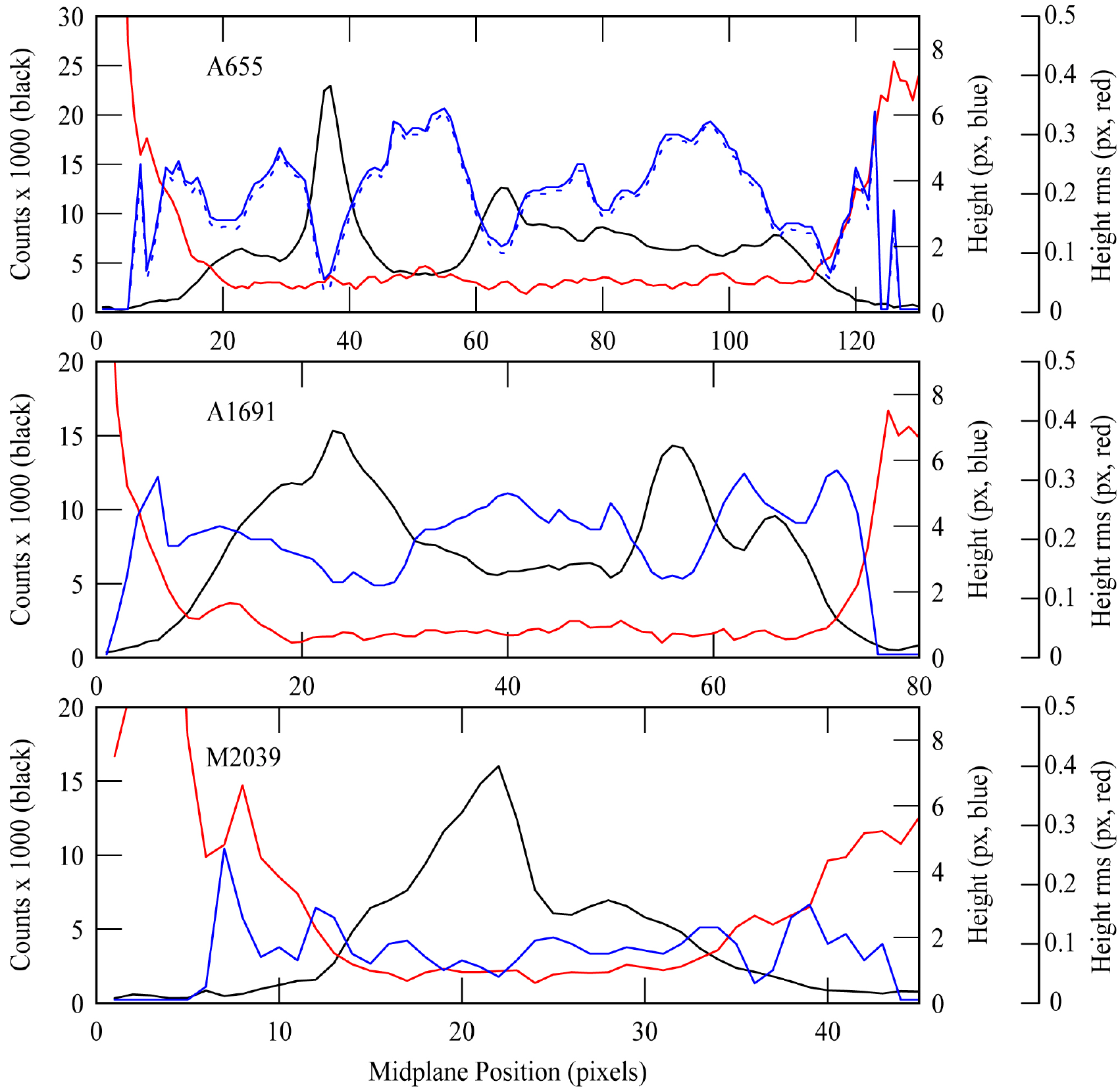}
\caption{Results for the three clumpy galaxies in Figure \ref{3clumpies-contours}.
In each panel,
the black curve shows the midplane F814W intensity measured along the major axis
of the galaxy, the blue curve shows the fitted scale height for each major axis
position, and the red curve shows the rms error between the fitted solution and the
observed scan. In the main part of the disk, there is an anti-correlation between the
midplane intensity and the fitted height that is modeled in Section \ref{model}
as the result of a superposition of a thick
and a thin disk component with variable relative intensities.  The dotted blue curve
for A655 shows the fitted scale heights when a clumpy line spread function is used
to deconvolve the vertical profiles.}
\label{frontier0655_1691_2039}
\end{figure}

Figure \ref{frontier0655_1691_2039} shows for the three clumpy galaxies in Figure
\ref{3clumpies-contours}: the midplane intensity as a function of position along the
midplane (black curves), the fitted height from perpendicular scans, also as a function
of midplane position (blue curves), and the rms between the fitted ${\rm sech}^{2}$
function and the perpendicular scan (red curves). The clumpy structure in these
galaxies can be matched to the peaks and valleys in the intensity scans. The scale for
intensity is counts$\times1000$ as indicated on the left-hand axis. The scales for
height and rms in pixels are on the right-hand axes. In the figure, the scale heights,
$H$, range from $\sim2$ pixels for M2039, to $\sim4$ pixels for A1691, to $\sim6$
pixels for A655.

\begin{figure}
\includegraphics[scale=0.5]{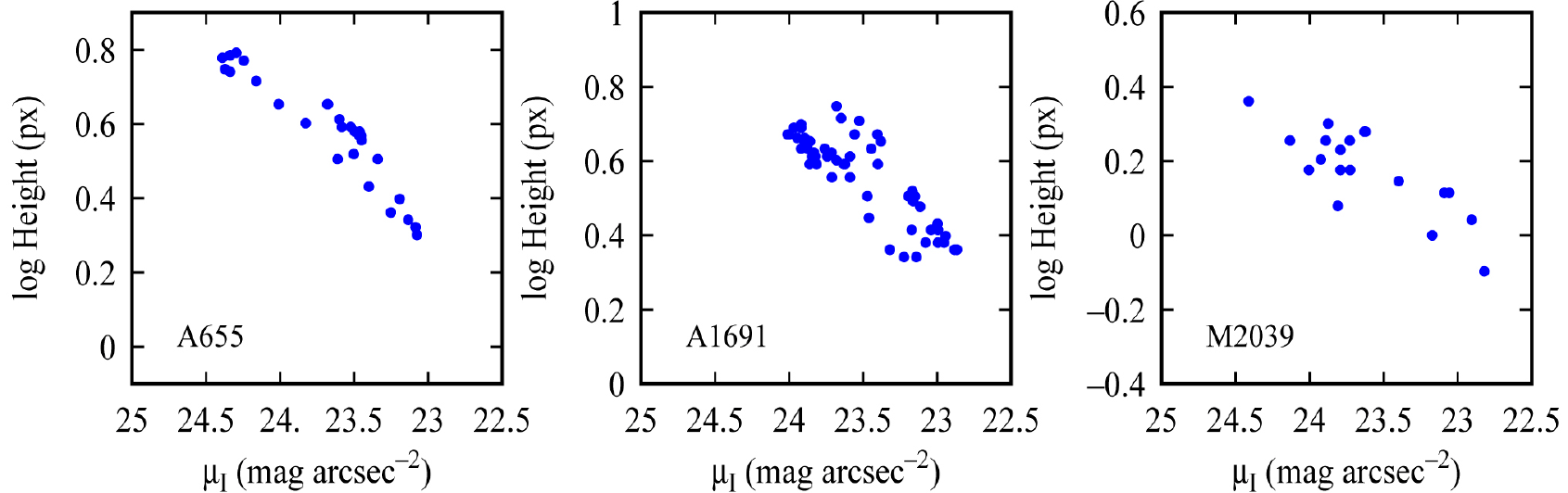}
\caption{The fitted scale heights, $H$, are plotted versus the midplane
intensities measured in magnitudes arcsec$^{-2}$ in the I band for the three clumpy
galaxies in Figure \ref{3clumpies-contours}.  The anti-correlation
between the height and the intensity is clear for these galaxies.}
\label{frontier0655_1691_2039_H_vs_inten}
\end{figure}

The top left panel of Figure \ref{frontier0655_1691_2039} has a dotted blue curve that
is the solution to the scale height when the clumpy line spread function (CLSF) is used
(cf. Sect. \ref{sect:lsf}). The model clump in the CLSF has a uniform brightness
between 5 and 11 pixels from the center of the PSF and an amplitude 6 times the
brightness of the line intensity.  These were the parameters that gave the maximum
deviation between the CLSF and the LSF in Section \ref{sect:lsf}. The dotted curve
shows that the fitted scale height decreases by only $\sim1$\% for deconvolution with
the CLSF compared to the LSF. This is a small change because the spacings between the
main clumps are usually well resolved by the PSF.

Figure \ref{frontier0655_1691_2039} reveals a pattern where regions of high intensity
correspond to relatively low scale heights, and vice versa. This pattern is shown
better in Figure \ref{frontier0655_1691_2039_H_vs_inten}, which plots, for the same
three galaxies, the log of the height in pixels versus the I-band surface brightness,
$\mu_{\rm I}$, in magnitudes per arcsec$^2$. The correlation has slopes of $d\log
H/d\mu_{\rm I}=0.32\pm0.04$, $0.28\pm0.04$, and $0.21\pm0.06$ for A655, A1691, and
M2039, respectively. Converting these magnitudes to intensity counts, $C_{\rm I}$, the
slopes are $d\log H / d\log C_{\rm I}=-0.79\pm0.10$, $-0.70\pm0.11$, and
$-0.53\pm0.16$, respectively. The uncertainties in the slopes were determined from the
Student-t distribution.

\begin{figure}
\includegraphics[scale=0.5]{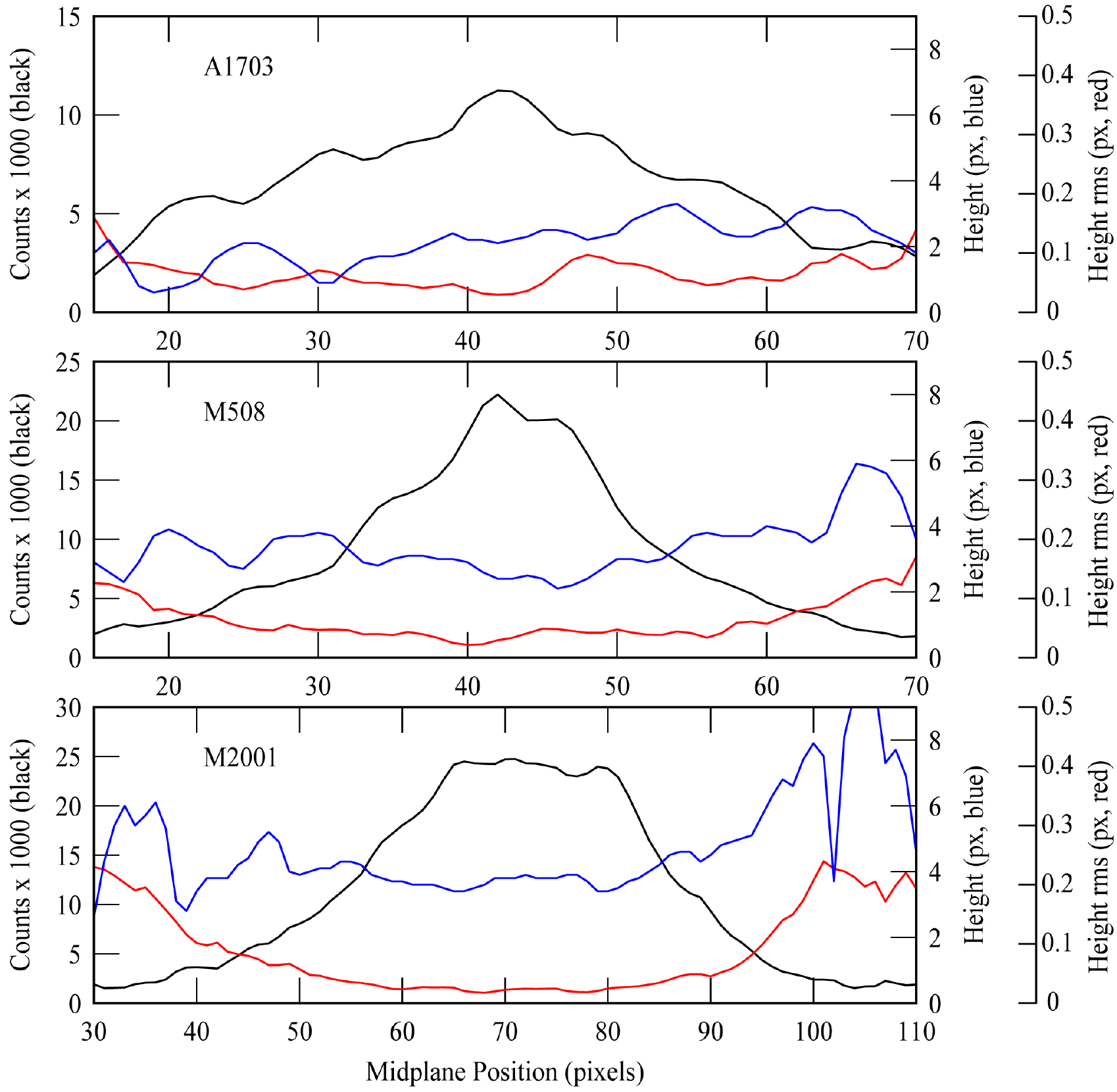}
\caption{Results for the three spiral galaxies in Figure \ref{3spirals-contours}.
Black curves are the midplane intensities, blue curves are the
fitted scale heights, and red curves are the rms errors, as in Figure
\ref{frontier0655_1691_2039}. There is very little anti-correlation between the
midplane intensity and the fitted height for spiral galaxies because
the thin and thick disks are both highly evolved.} \label{frontier1703_508_2001}
\end{figure}

\begin{figure}
\includegraphics[scale=0.5]{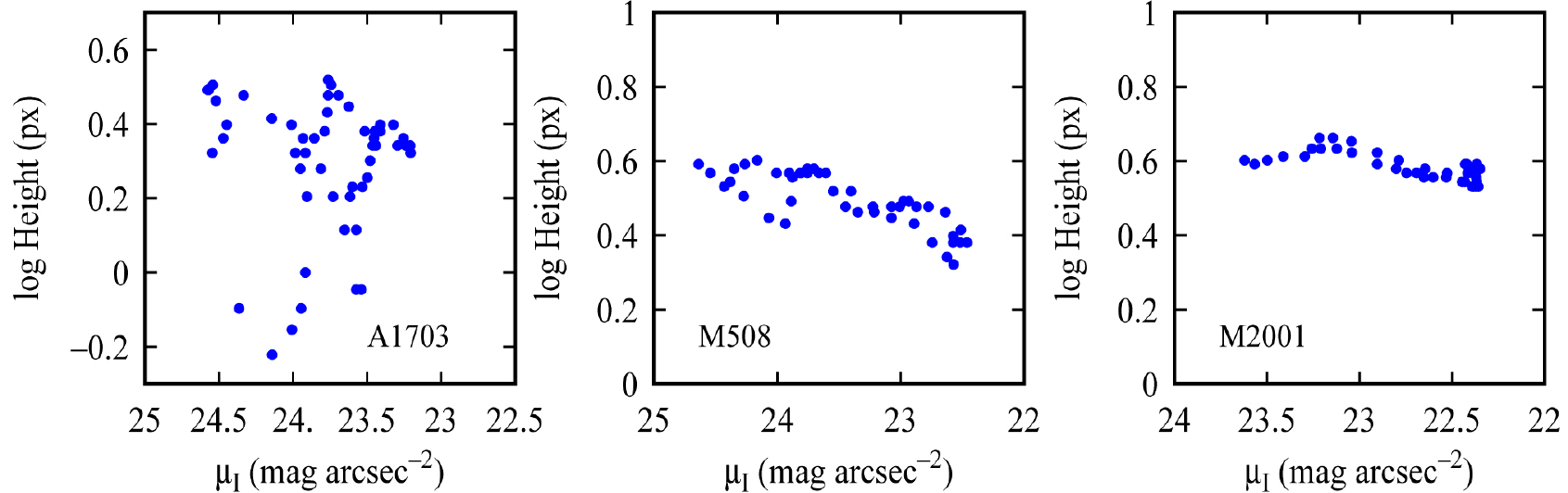}
\caption{The fitted scale heights, $H$, are plotted versus the midplane
intensities measured in magnitudes arcsec$^{-2}$ in the I band for the three spiral
galaxies in Figure \ref{3spirals-contours}.}
\label{frontier1703_508_2001_H_vs_inten}
\end{figure}

Figure \ref{frontier1703_508_2001} shows the same midplane scans and fits for the three
spiral galaxies in Figure \ref{3spirals-contours}, and Figure
\ref{frontier1703_508_2001_H_vs_inten} shows the heights versus the midplane surface
brightnesses for these galaxies.  The anti-correlation is not as strong for spirals as
it is for the clumpy types.

\begin{figure}
\includegraphics[scale=0.8]{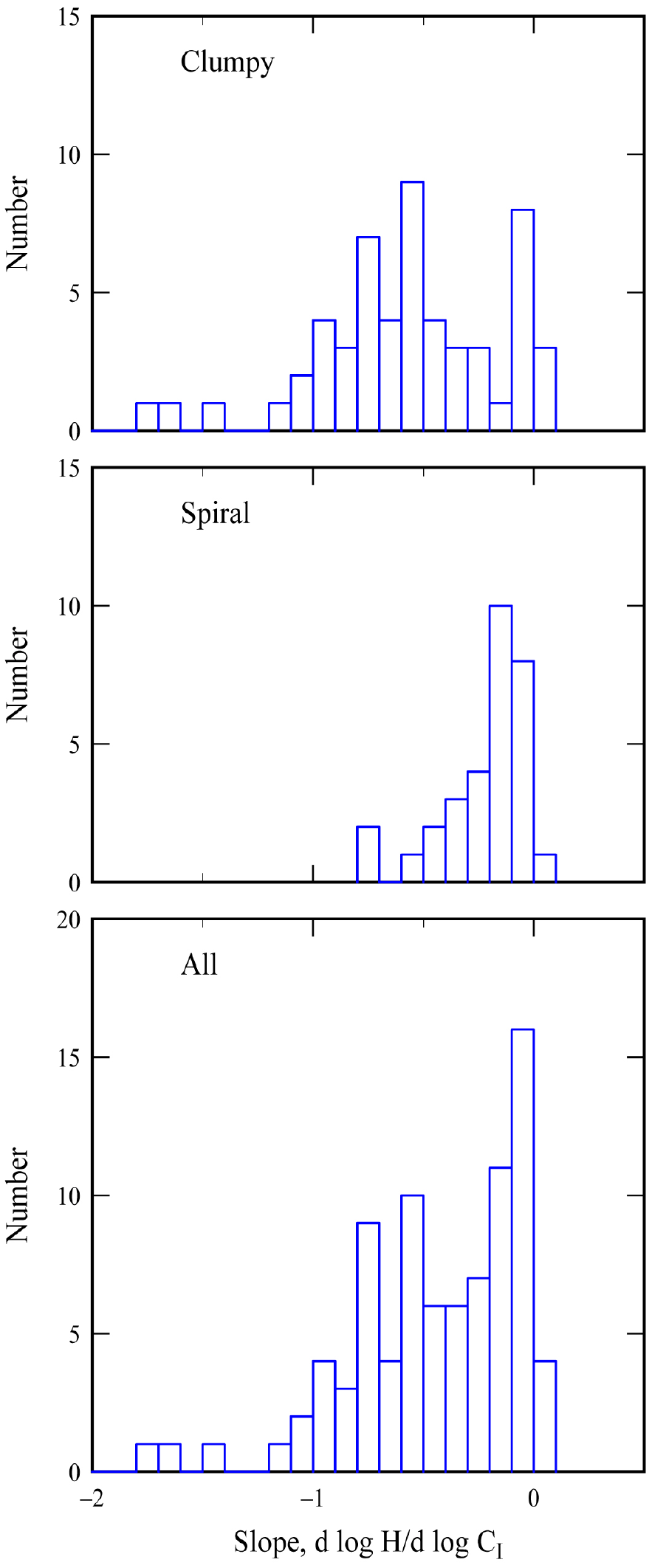}
\caption{Histograms are shown for the slope of the anti-correlation
between the fitted scale height and the midplane intensity.  The scale height is $H$ and
the I-band intensity measured in counts is $C_{\rm I}$. The strong
anti-correlation for the clumpy galaxies shows up here as a significant
shift to negative values for the correlation slope. The shift is less for
spiral galaxies. } \label{frontier_histogram2}
\end{figure}

Height-intensity anti-correlations like this were fitted for all of the galaxies in our
sample using plots of the log of the height in pixels versus the log of the intensity
in counts. Figure \ref{frontier_histogram2} shows histograms of the slopes in these
anti-correlations. The bottom histogram is for the spiral and clumpy galaxies combined,
the middle histogram is for the spirals, and the top histogram is for the clumpy
galaxies. The average slope for the all of the galaxies is $<d \log H / d\log C_{\rm
I}>= -0.38\pm0.73$; for the clumpy galaxies it is $-0.47\pm0.84$, and for the spirals
it is $-0.17\pm0.30$. The spirals have a significantly weaker height-intensity
anti-correlation than the clumpy galaxies.

\begin{figure}
\includegraphics[scale=0.4]{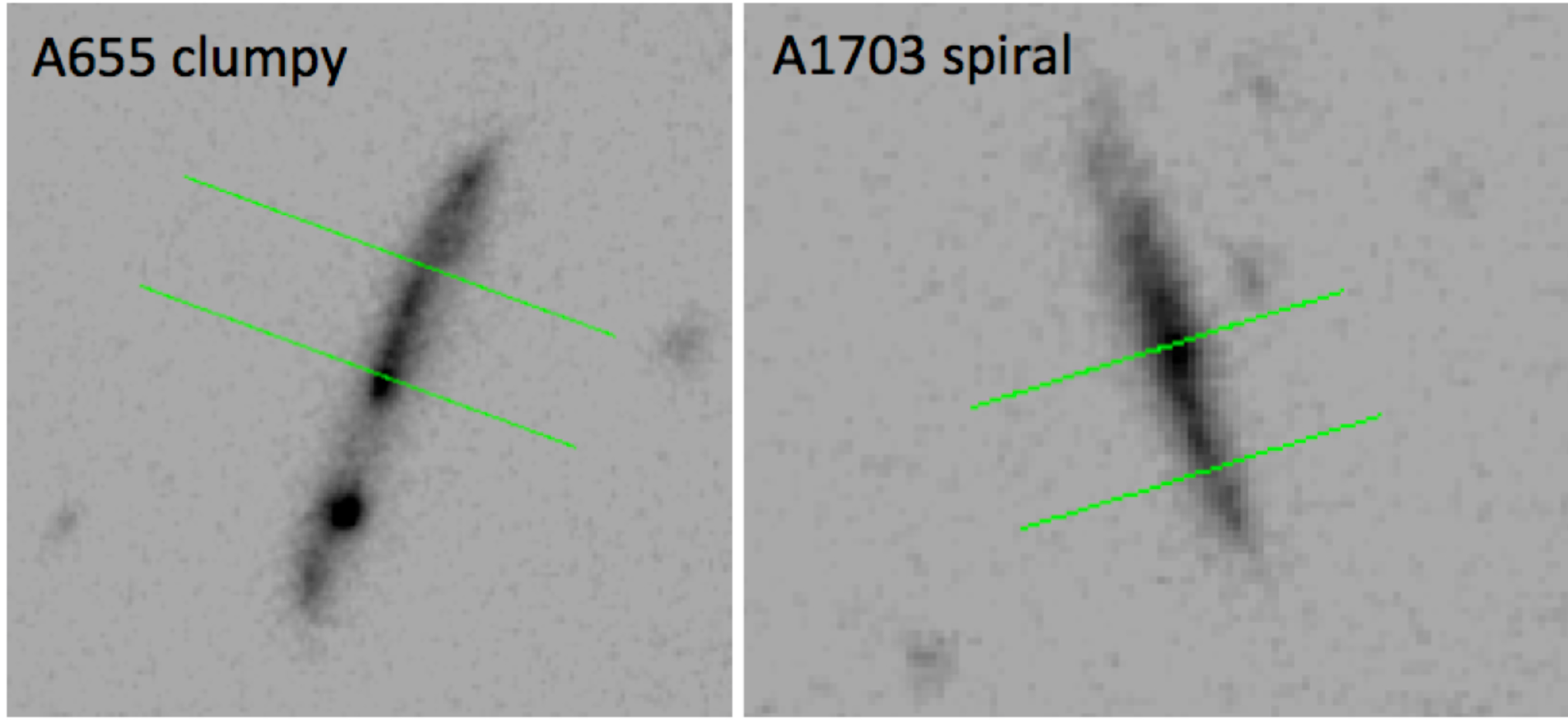}
\caption{Two galaxies from Figures \ref{3clumpies-contours}
and \ref{3spirals-contours} with perpendicular cut positions
used to illustrate the difference between perpendicular profiles in Figure
\ref{frontier0655_1703_profiles}. } \label{A655_1703cut}
\end{figure}

\begin{figure}
\includegraphics[scale=0.5]{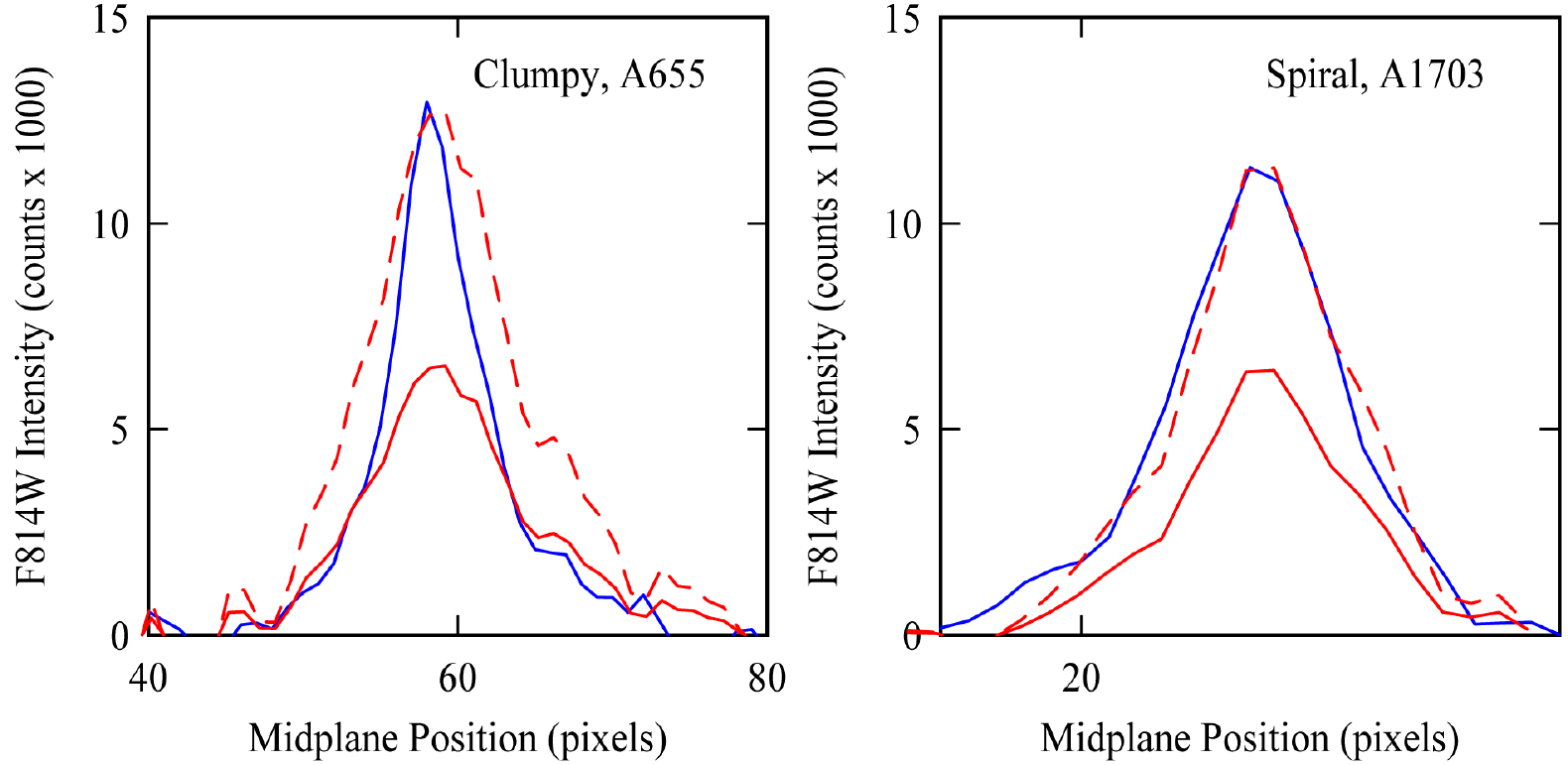}
\caption{Perpendicular intensity profiles for the cuts indicated in
Figure \ref{A655_1703cut} showing how the clump profiles
are thinner when they are higher, but the spiral profiles are about the
same width. The raw profiles are plotted as solid curves with the stronger
profile in blue. The dashed red curve is the same as the solid red curve
but multiplied by a constant factor to give the same peak.
} \label{frontier0655_1703_profiles}
\end{figure}

An example of this difference between galaxies that have an anti-correlation and those
that do not is shown in Figures \ref{A655_1703cut} and
\ref{frontier0655_1703_profiles}. The left-hand panel of Figure
\ref{frontier0655_1703_profiles} has two perpendicular intensity scans for a clumpy
galaxy (galaxy A655 in Fig. \ref{3clumpies-contours}), one from a bright clump and
another from a faint interclump region shown as solid curves, and the right-hand panel
has two perpendicular intensity scans for a spiral galaxy (galaxy A1703 in Fig.
\ref{3spirals-contours}), one from the bulge and another from the disk. The regions
chosen are shown in Figure \ref{A655_1703cut}. In each panel of Figure
\ref{frontier0655_1703_profiles}, the dashed red curve is the same as the solid red
curve but stretched upward to match the peak of the solid blue curve. For the clumpy
galaxy, the bright region with the higher peak (blue curve) is clearly narrower than
the faint region (red curve), while for the spiral galaxy, both regions have about the
same profile widths.

This difference in height-intensity correlation for spiral and clumpy galaxies is also
evident directly from the contours in Figure \ref{3clumpies-contours}, as mentioned
previously. The contours around the clumpy galaxies are fairly straight on each side of
the clumps, whereas the contours bow out to a $V$-shape around the bright parts of the
spiral. For the clumpy galaxies, the straight contours mean that the intensity at a
certain height above the midplane is independent of the intensity in the midplane, and
for this to be true, the scale height has to be smaller in the regions of higher
intensity. For the spirals, the greater intensity at fixed height above the brighter
regions corresponds to an increase in the intensity of the whole vertical profile with
an approximately fixed profile width.

\subsection{Model for the anti-correlation}
\label{model}

The anti-correlation for clumpy galaxies, and the average value of its slope,
$-0.47\pm0.84$, make sense if there are two components in each galaxy, a bright thin
disk and a faint thick disk, which blend together into our single ${\rm sech}^{2}$
fitting function at the relatively poor resolution of the survey.  Consider what
happens as the thin component gets brighter.  Let $A_{\rm thin}\exp(-z/H_{\rm thin})$
be an approximation for the thin component at $z>H_{\rm thin}$ and let $A_{\rm
thick}\exp(-z/H_{\rm thick})$ be an approximation for the thick component off the
midplane. When the thin component is faint, the profile is dominated by the thick
component and the measured thickness is $H_{\rm thick}$. This remains true until the
brightness of the thin component at a height equal to the scale height of the thick
component becomes equal to the brightness of the thick component there,
\begin{equation}
A_{\rm thin}e^{-H_{\rm thick}/H_{\rm thin}}\sim A_{\rm thick}e^{-1}.
\end{equation}
Then the main part of the thick disk is dominated in brightness by the thin component
and the measured scale height is the thin component value. Thus there is a transition
from the thick component scale height to the thin component scale height as the thin
component brightness increases from zero to
\begin{equation}
A_{\rm thin}=A_{\rm thick}e^{(H_{\rm thick}/H_{\rm thin}-1)}.
\end{equation}
The logarithmic derivative corresponding to this transition is approximately
\begin{equation}
{{d \log H}\over{ d \log A}} \sim -{{\ln (H_{\rm thick}/H_{\rm thin})}\over
{H_{\rm thick}/H_{\rm thin}-1}}.
\end{equation}
For $H_{\rm thick}/H_{\rm thin}$ equal to typical values of 3 to 5 \citep{comeron11},
this logarithmic derivative ranges between -0.55 and -0.40, as observed in Figure
\ref{frontier_histogram2}.

\begin{figure}
\includegraphics[scale=0.7]{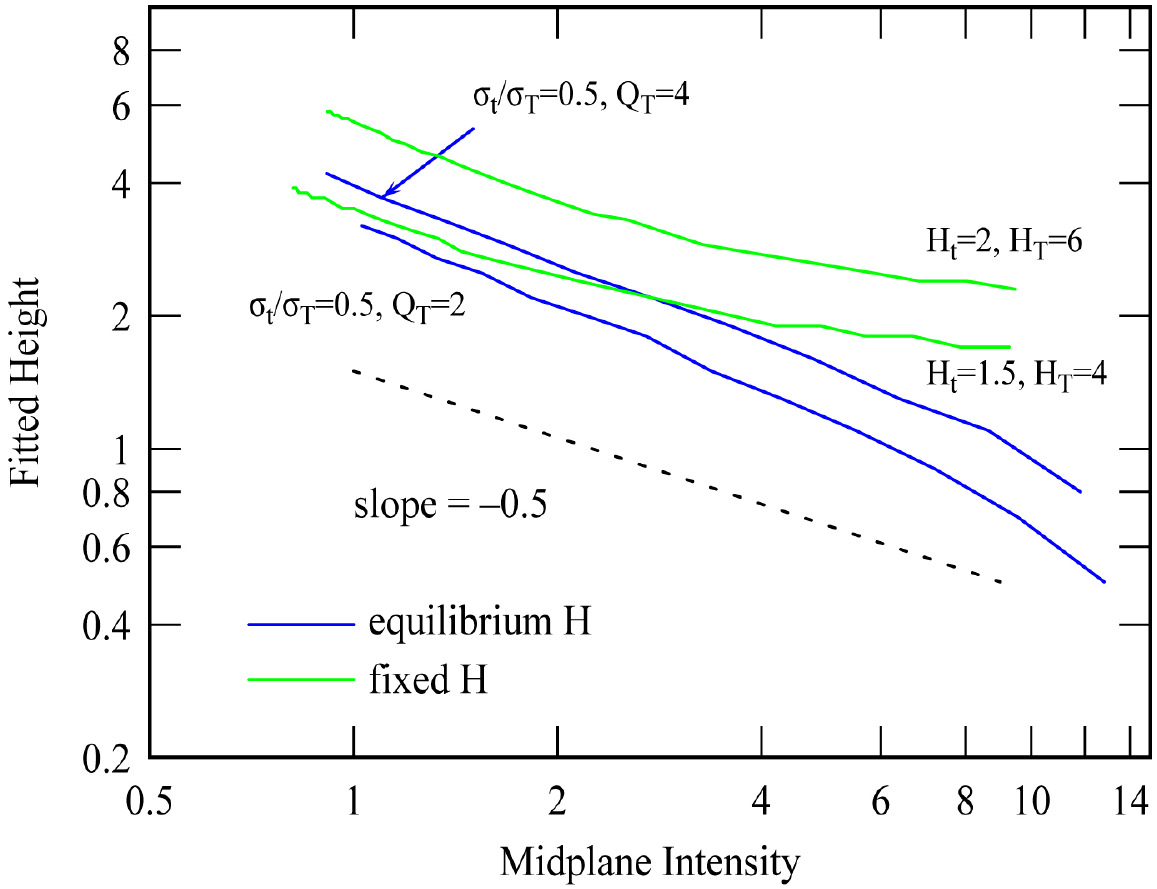}
\caption{Two models for the anti-correlation between the height that
is determined by fitting a two-component disk with a single ${\rm sech}^{2}$ function
and the midplane intensity.  The green curves have fixed scale heights for the
two components and a variable intensity for the thin component, while the
blue curves have two components in vertical equilibrium with a variable
mass for the thin component. Both models have a slope of about $-0.5$,
which is what we observe for the clumpy galaxies. The midplane intensity is
from the sum of the two components, as would be observed for a real galaxy.}
\label{frontier_vertical2}
\end{figure}

Two detailed models for this anti-correlation were also considered. The first makes a
two component disk with fixed scale heights for each component and a fixed central
brightness for the thick component. Then it varies the central brightness of the thin
component, blurs the combined disk with our LSF, and stores the result as a mock
observation. This observation is then fed into our main deconvolution program to
recover the input disk as fitted by a single ${\rm sech}^{2}$ function.  The resulting
thickness of the fitted composite disk is plotted versus the total measured intensity
of the midplane (from the blurred sum of the thin and thick disks) as a green line in
Figure \ref{frontier_vertical2} for two cases, one with $H_{\rm thin}=2$ and $H_{\rm
thick}=6$, and another with $H_{\rm thin}=1.5$ and $H_{\rm thick}=4$. Both have fitted
heights at low intensity that are about equal to the thick disk scale height, and
fitted heights at high intensity that are about equal to the thin disk scale height.
The anti-correlation has a slope of $\sim-0.5$, as shown by the fiducial dotted line.
This is similar to what we observe in the real galaxies.

The second model makes a two-component disk that is in hydrostatic equilibrium
following equations in \cite{narayan02}, but without the dark matter component. We
assume a ratio of velocity dispersions for the thin and thick disks equal to 0.5, a
Toomre-Q value for the thick disk equal to 2 and 4 in two cases, and a variable ratio
of the surface density of thin disk to thick disk.  As the thin disk component
increases in mass relative to the thick disk, the thick disk scale height decreases
because of the larger gravitational attraction to the midplane. Thus there are two
effects causing an anti-correlation in this case. The results are shown as blue curves
in Figure \ref{frontier_vertical2}. The slope is also about $-0.5$.

The anti-correlation slope for the model with fixed scale heights is about the same as
the slope for the hydrostatic model with variable scale heights because the brightening
effect of the thin disk dominates the correlation in both cases. The shrinking thick
disk in the equilibrium case cannot be seen beneath the brightening of the thin disk.
If the thin disk were darker, adding mass but not light, or if the thick disk could be
observed further from the midplane where the thin disk is very weak, then the shrinking
of the thick disk with increasing thin disk mass should be observable.  This is
evidently an observation well suited for the James Webb Space Telescope, with its
heightened sensitivity in the near-infrared.

An anti-correlation between scale height and midplane surface brightness should be
expected in gas-dominated galaxies because star formation in the relatively thin
component can be much brighter than the older stars in the thick component. A more
evolved galaxy with less star formation should have less of an anti-correlation because
the thin disk is less prominent compared to the thick disk. Presumably this is why the
spiral galaxies in our survey have only weak anti-correlations between scale height and
midplane surface brightness (Fig. \ref{frontier_histogram2}).  Spiral galaxies are more
evolved than clumpy galaxies \citep{elmegreen14}.

\subsection{Color Differentials over Height}

The scale heights were measured for all three filters, B, V and I, but the B images
were faint and had the largest rms values for the fits.  Here we compare the V and I
scale heights to search for vertical color gradients. The V band intensity at one scale
height in the I band is
\begin{equation}
I_{\rm V}=I_{\rm V,0}\left({{2}\over{e^{H_{\rm I}/H_{\rm V}}+e^{-H_{\rm I}/H_{\rm V}}}}\right)^2,
\end{equation}
and the I band intensity at one scale height in the I band is
\begin{equation}
I_{\rm I}=I_{\rm I,0}\left({{2}\over{e^{1}+e^{-1}}}\right)^2.
\end{equation}
The $V-I$ color differential from the midplane to one scale height in I band is
\begin{equation}
\Delta (V-I) = (V-I)_{H_I}-(V-I)_0 = -5\log\left({{e^{1}+e^{-1}}\over{e^{X}+e^{-X}}}\right)
\end{equation}
where $X=H_{\rm I}/H_{\rm V}$.

\begin{figure}
\includegraphics[scale=0.8]{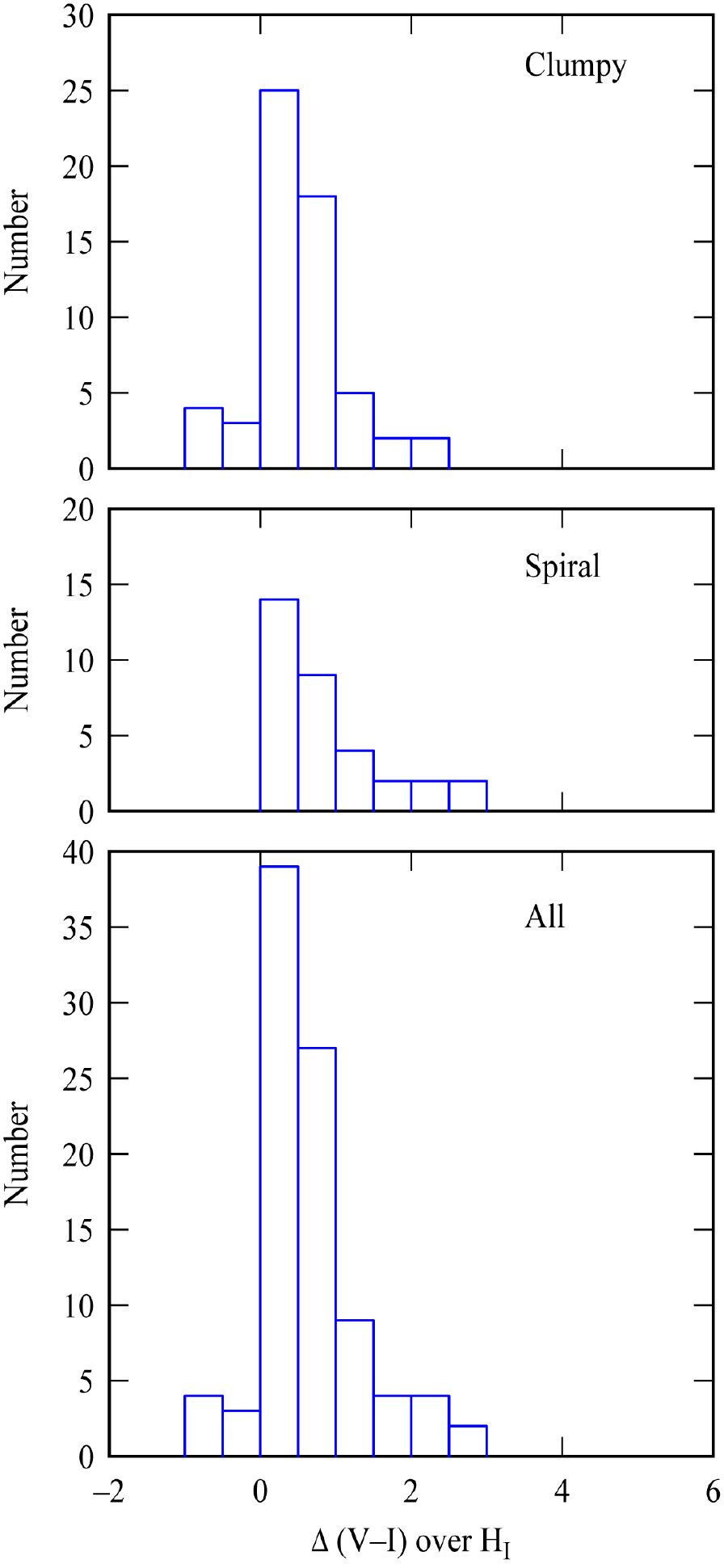}
\caption{Histogram of the color differences between the midplanes of the galaxies and the
positions at one scale height in I band above the midplanes. }
\label{frontier_histogramcolor}
\end{figure}

The distribution functions for $\Delta(V-I)$ among all the galaxies are shown in Figure
\ref{frontier_histogramcolor}, divided into galaxy type. Although the distribution
functions are broad, there is a slight offset toward redder $V-I$ at one I-band scale
height compared to the midplane. The average $V-I$ reddenings are $0.63\pm0.68$,
$0.89\pm0.76$, and $0.48\pm0.59$ for all galaxies, spirals, and clumpy galaxies,
respectively.

\begin{figure}
\includegraphics[scale=0.5]{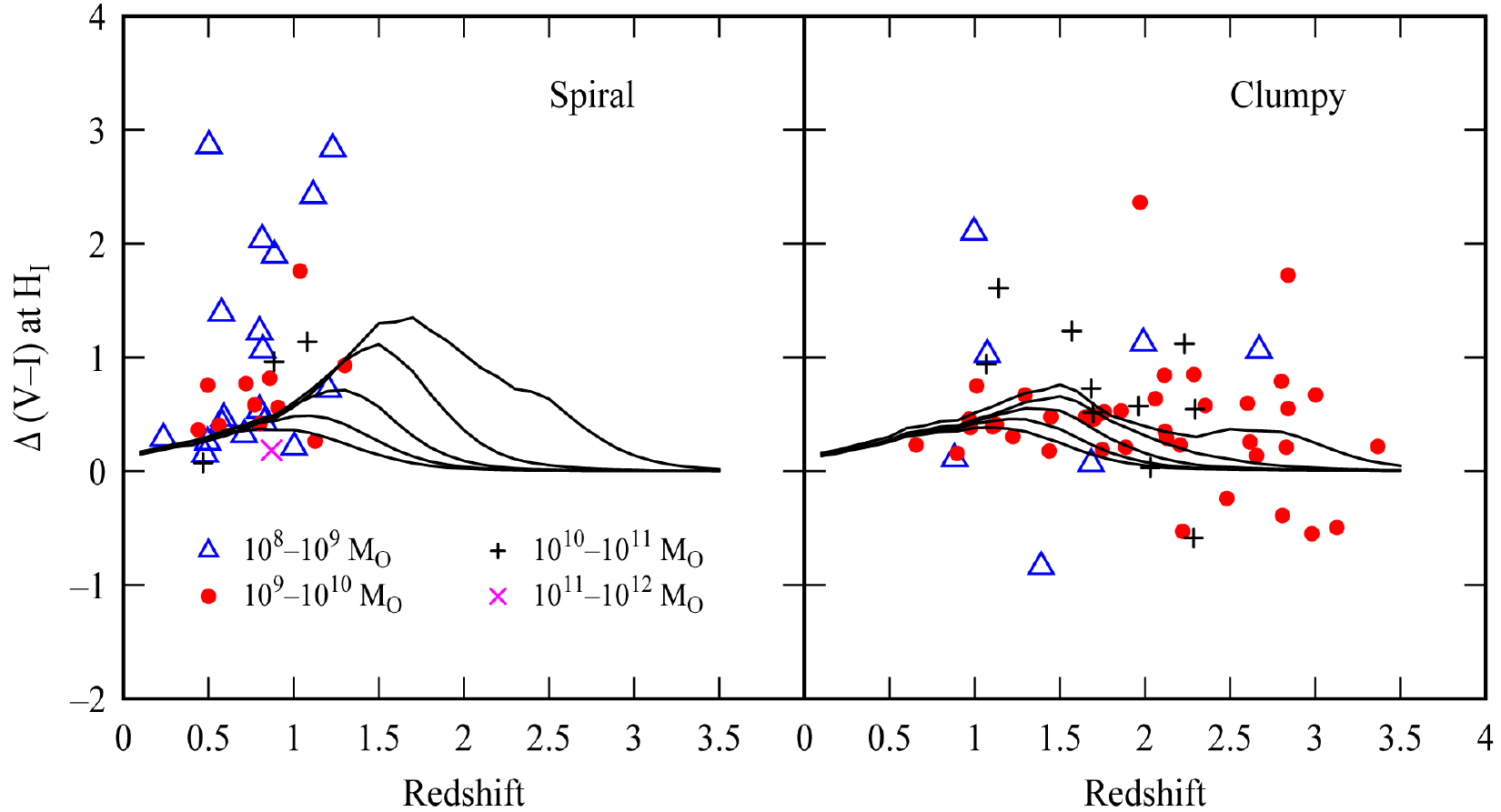}
\caption{The color differences between the midplane and one scale height in I band
are plotted versus the redshifts for different mass ranges indicated by different
symbols. The curves are models based on star formation and stellar evolution
in the thick and thin disk components. On the right, the curves have decay times, $\tau$
in equation \ref{eq:11}, that increase from 0.2 Gyr for the upper curve to 1 Gyr for the
lower curve.  The model assumes that the final thick disk mass equals the final
thin disk mass. The curves are a reasonable fit to the red dots, which have an intermediate mass, as
indicated on the left. The curves on the left have the same range of $\tau$ but
assume that the final thick disk mass is 3 times the final
thin disk mass. They reach higher color differences but
are not an ideal fit to the data.} \label{frontier_color_vs_z_different_masses2}
\end{figure}

Reddening of the starlight with distance from the midplane is another indication that
there is a range of stellar populations including a thin young component and a thick
old component. The corresponding age difference for a given reddening value depends on
the redshift because of bandshifting. Figure
\ref{frontier_color_vs_z_different_masses2} shows the $V-I$ color differences at the I
band scale height as a function of redshift with different symbols representing
different mass ranges, $10^7-10^8\;M_\odot$, $10^8-10^9\;M_\odot$,
$10^9-10^{10}\;M_\odot$, and $10^{10}-10^{11}\;M_\odot$. The galaxies are divided into
spiral and clumpy types. The spirals only occur at low redshifts, less than $\sim1.5$,
as found previously \citep{elmegreen14}, whereas the clumpy types extend to higher
redshifts.

The curves on the panel for the clumpy galaxies in Figure
\ref{frontier_color_vs_z_different_masses2} are models using \cite{bruzual03} stellar
population spectra that were redshifted and integrated over the HST ACS filters used
here. They include absorption from the intervening Lyman $\alpha$ forest
\citep{madau95} but no dust absorption (dust absorption in the thin disk component
decreases the thick-minus-thin disk color difference, possibly making it negative). The
different curves are for different values of $\tau$ in a star formation history that is
modeled as
\begin{equation}
{\rm SFR}_{\rm thick}=Ae^{-t/\tau} \;\;;\;\;{\rm SFR}_{\rm thin}=B(1-e^{-t/\tau}).
\end{equation}
These formulae assume the thick disk forms first with an exponentially decaying rate,
and the thin disk forms second with an exponentially increasing rate to a constant
value. The coefficients $A$ and $B$ determine the relative star formation rates and
therefore the relative final masses in the two components. Because what we need for
comparison with the observations is the color gradient, we choose $B=1$ and
\begin{equation}
A={{T-\tau(1-e^{-T/\tau})}\over{\tau(1-e^{-T/\tau})}},
\label{eq:11}
\end{equation}
which gives the thick and thin disk an equal mass after a Hubble time, $T$. The assumed
values of $\tau$ are 0.2 Gyr, 0.4 Gyr, up to 1 Gyr, in equal steps. Figure
\ref{frontier_color_vs_z_different_masses2} shows a reasonable agreement between the
model and the observations of clumpy galaxies in the mass range from
$10^9-10^{10}\;M_\odot$. The low-mass spiral galaxies (on the left in Figure
\ref{frontier_color_vs_z_different_masses2}) have higher color differentials that we
were not able to reproduce with this method.  The curves in the left panel assume that
the final thick disk mass is three times more than the final thin disk mass. This
increases the color difference. Larger or smaller mass thick disks do not have even
higher color differentials because in the first case the midplane starts to take on the
color of the thick disk, decreasing the differential compared to the color at one scale
height, and in the second case the disk at one scale height becomes dominated by the
thin disk component, decreasing the color differential again.

\section{Correlations with Galaxy Mass and Redshift}

\begin{figure}
\includegraphics[scale=0.5]{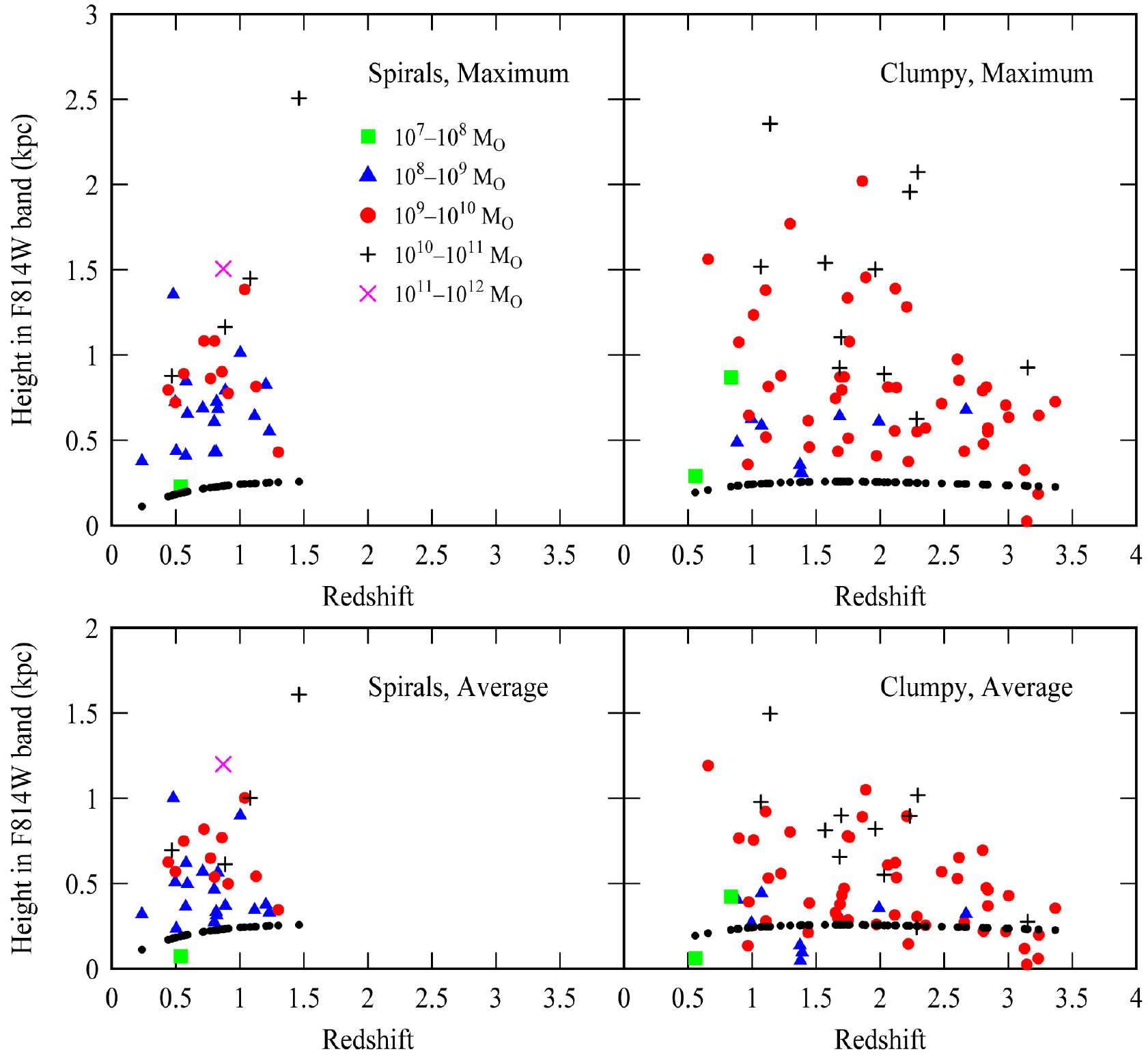}
\caption{The scale heights in I band are shown versus the redshifts for spiral
galaxies on the left and clumpy galaxies on the right. Galaxy mass ranges are
indicated by the different symbols.  The black dots at the bottom of each panel
are the sizes of a single pixel at that redshift (one dot per galaxy). The top panels
are the maximum heights measured for each galaxy, considered to be representative of the
thick disk unblended with the thin disk, while the lower panels are the average
heights for all of the single-component fits in each galaxy.}
\label{frontier_h_vs_z_different_masses2}
\end{figure}

Figure \ref{frontier_h_vs_z_different_masses2} shows the fitted scale heights in I-band
converted to physical sizes in kpc using redshifts from \cite{castellano16}, plotted
versus the galaxy redshifts with different symbols representing different intervals of
galaxy mass. The black dots represent the size of a single pixel at the redshift of a
corresponding galaxy. The bottom two panels plot the average height in each galaxy,
averaged over all of the pixels along the major axis where there is a good fit with a
low rms. These good-fit regions were determined for each galaxy by eye using plots like
Figure \ref{frontier0655_1691_2039}. Typically, a range of pixels on each side of the
nucleus was selected for the spiral galaxies in order to avoid the bulge. For the
clumpy galaxies, the selected region usually extended through the whole disk. This
averaging procedure means that the plotted values are between the high and low values
of the fitted heights in the anti-correlation discussed in Section \ref{anti}. To get a
value more representative of the thick disk, the top panels plot the maximum height
from all of the fitted values in the same pixel range. This maximum usually occurs
where the midplane intensity is small. According to the models in Section \ref{model},
it should represent the thick disk scale height better than the average.

Figure \ref{frontier_h_vs_z_different_masses2} suggests that the thickest clumpy
galaxies in the mass range $10^9-10^{10}\;M_\odot$ (red points) tend to get thicker at
decreasing redshift. That is, the top of the distribution of red points increases
toward decreasing redshift. The thickest clumpy galaxies are also thicker than the
spiral galaxies at the same redshift.  To be specific, we determined average heights
and maximum heights for clumpy galaxies with masses between $10^{9.5}$ and $10^{10.5}$
and for three redshift bins: 0.5-1.5, 1.5-2.5, and 2.5-3.5. The average heights in kpc
are, respectively, $1.03\pm0.25$, $0.63\pm0.24$ and $0.32\pm0.21$. The maximum heights
are $1.61\pm0.39$, $1.06\pm0.43$, and $0.60\pm0.27$. In comparison, spiral galaxies in
the same mass range and for redshifts between 0.5 and 1.5 have an average height of
$0.80\pm0.14$ and a maximum height of $1.13\pm0.17$.

\begin{figure}
\includegraphics[scale=0.5]{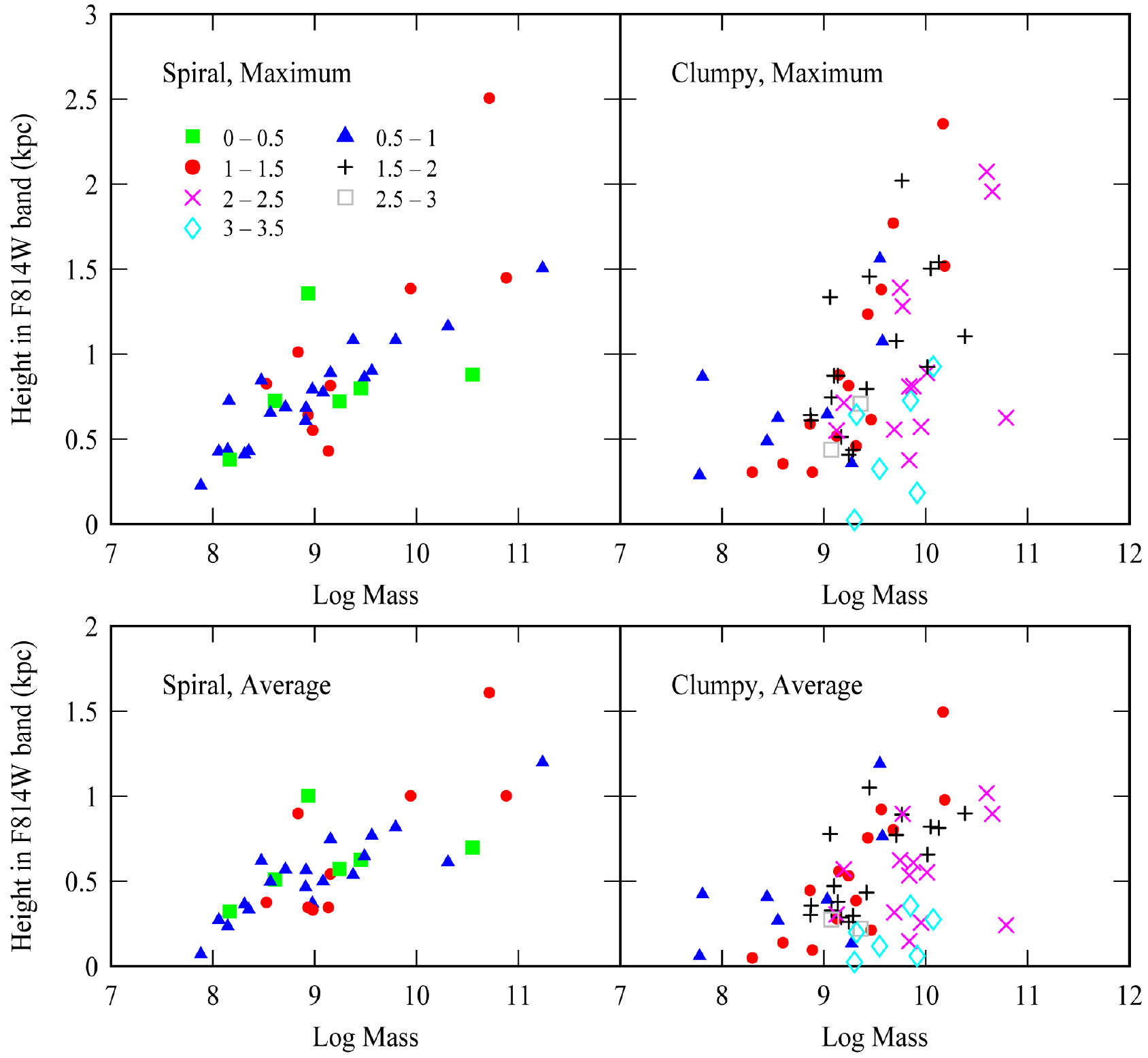}
\caption{The height in I band is shown versus the galaxy mass with different symbols
representing different redshift ranges as indicated in the left-hand panel. Spirals are on the left and clumpy galaxies
are on the right. The maximum heights for each galaxy are in the top panels and the
average heights for each galaxy are in the bottom panels. }
\label{frontier_h_vs_m_different_redshifts}
\end{figure}

\begin{figure}
\includegraphics[scale=0.9]{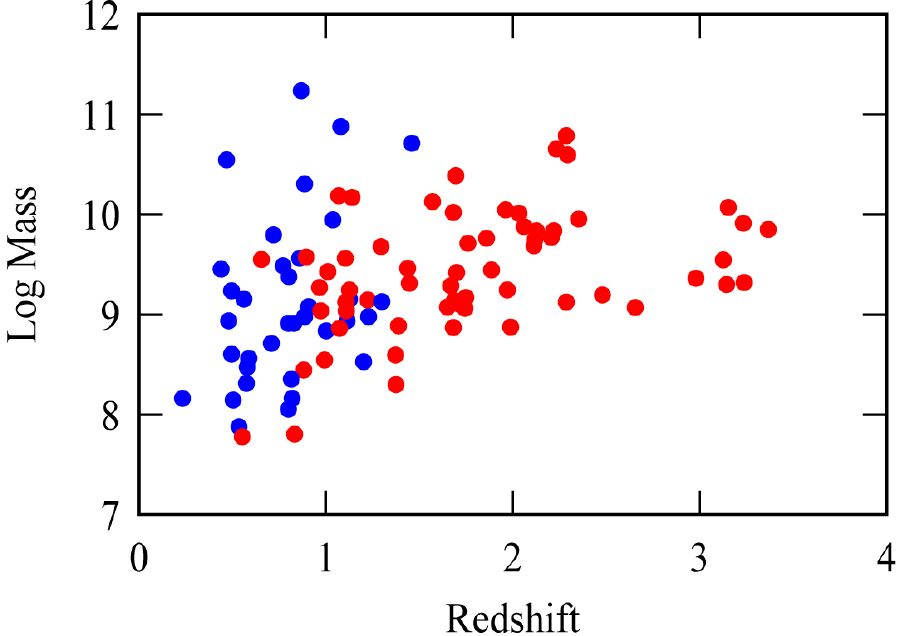}
\caption{Galaxy mass versus redshift with blue points for the spirals and
red points for the clumpy galaxies.} \label{frontier_m_vs_redshift}
\end{figure}

This redshift correlation is probably the result of a mass correlation shown in Figure
\ref{frontier_h_vs_m_different_redshifts}, which plots the average and maximum
thicknesses for spiral and clumpy types as a function of mass with different symbols
representing different redshift ranges, as indicated in the top left panel. There is a
clear correlation between height and mass for all redshifts.  Because the upper range
of mass increases with decreasing redshift (Fig. \ref{frontier_m_vs_redshift}), the
height-mass correlation probably explains the height-redshift correlation in Figure
\ref{frontier_h_vs_z_different_masses2}.

\begin{figure}
\includegraphics[scale=0.5]{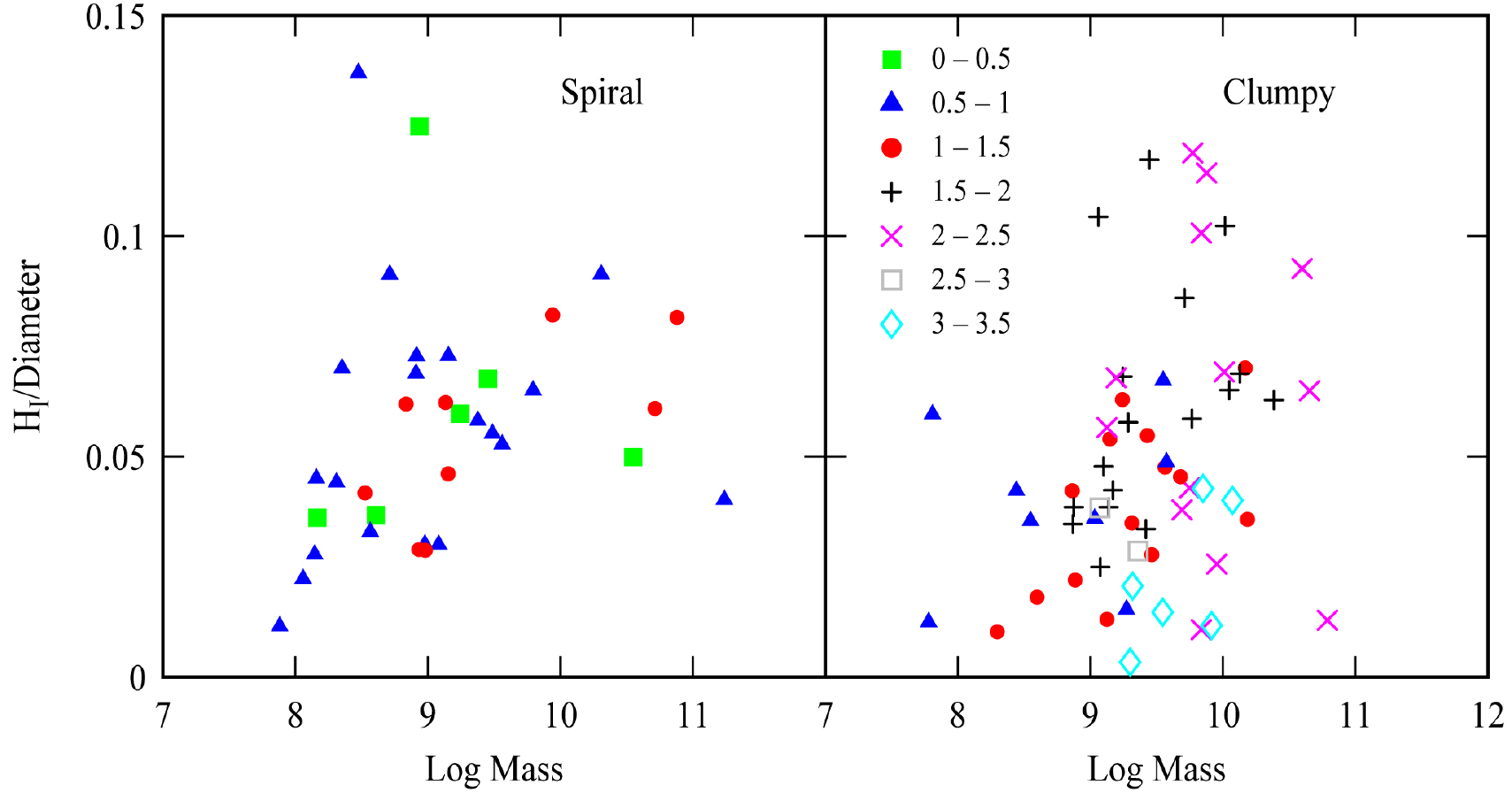}
\caption{The ratio of disk height to disk diameter versus galaxy mass
with different symbols representing different redshifts, as indicated in the right-hand
panel.} \label{frontier_hoverR_vs_m_different_redshifts2}
\end{figure}

Figure \ref{frontier_hoverR_vs_m_different_redshifts2} shows the ratio of the average
scale height for each galaxy to the galaxy radius determined from the extent of the
disk in the I-band, down to about the $3\sigma$ contour. Different symbols are for
different redshift ranges. The upper limit to this ratio increases with mass but is
approximately the same for spiral and clumpy types and independent of redshift.

\section{Conclusions}

The vertical scale heights of edge-on spiral and clumpy galaxies in two HST Frontier
Field Parallels were measured after deconvolution from the instrument point spread
function. The scale heights anti-correlate with midplane brightness for the clumpy
galaxies, suggesting two components, a bright thin disk and a faint thick disk, with
variations in the relative brightnesses of these components as a result of clumpy
midplane star formation.  Three models reproduced this anticorrelation. The vertical
profiles also reddened slightly with height, suggesting the same two components with a
difference in their formation times equal to a fraction of a Gyr. The scale heights
increase with galaxy mass for both galaxy types and for all observed redshifts.

The observations indicate that clumpy galaxies have thick disks with somewhat thinner
star formation components at redshifts out to at least $z=3$. Spiral galaxies also have
thick disks out to $z\sim1.5$, as measured directly, but the anti-correlation between
height and brightness is weaker than for clumpy types. This is presumably because the
thin disk is older and the star formation bursts are weaker for spirals than clumpy
types, and therefore the spiral thin disk has less contrast to its old thick disk.

{\it Acknowledgments} We gratefully acknowledge the National Science Foundation Grant
AST-1005024 to the Keck Northeast Astronomy Consortium REU Program and the Vassar
College Undergraduate Research Summer Institute. We thank the referee for useful
comments.

\end{document}